\documentclass[10pt,british,english,aps,superscriptaddress,preprintnumbers,amsmath,nofootinbib,amssymb,altaffilletter]{revtex4}
\usepackage[T1]{fontenc}
\usepackage[latin9]{inputenc}
\usepackage{xcolor}
\usepackage{babel}
\usepackage{amsmath}
\usepackage{amssymb}
\usepackage{graphicx}
\usepackage{esint}
\usepackage[unicode=true,pdfusetitle,
 bookmarks=true,bookmarksnumbered=false,bookmarksopen=false,
 breaklinks=false,pdfborder={0 0 1},backref=false,colorlinks=false]
 {hyperref}

\makeatletter
\@ifundefined{textcolor}{}
{%
 \definecolor{BLACK}{gray}{0}
 \definecolor{WHITE}{gray}{1}
 \definecolor{RED}{rgb}{1,0,0}
 \definecolor{GREEN}{rgb}{0,1,0}
 \definecolor{BLUE}{rgb}{0,0,1}
 \definecolor{CYAN}{cmyk}{1,0,0,0}
 \definecolor{MAGENTA}{cmyk}{0,1,0,0}
 \definecolor{YELLOW}{cmyk}{0,0,1,0}
}

\usepackage{babel}
\usepackage{babel}
\usepackage{babel}
\usepackage{babel}
\usepackage{babel}

\usepackage{babel}
\usepackage{breakurl}

\@ifundefined{textcolor}{}{%
 \definecolor{BLACK}{gray}{0}
 \definecolor{WHITE}{gray}{1}
 \definecolor{RED}{rgb}{1,0,0}
 \definecolor{GREEN}{rgb}{0,1,0}
 \definecolor{BLUE}{rgb}{0,0,1}
 \definecolor{CYAN}{cmyk}{1,0,0,0}
 \definecolor{MAGENTA}{cmyk}{0,1,0,0}
 \definecolor{YELLOW}{cmyk}{0,0,1,0}
}

\usepackage{babel}
\usepackage{epsfig}\usepackage{mathrsfs}\usepackage{euscript}

\renewcommand{\[}{\begin{equation}}
\renewcommand{\]}{\end{equation}}
\def\beq{\begin{equation}}
\def\eeq{\end{equation}}
\newcommand{\be}{\begin{eqnarray}}
\newcommand{\ee}{\end{eqnarray}}

\renewcommand{\texttt}{{}}

\def\bs{\begin{subequations}}
\def\es{\end{subequations}}

\def\Fc{\mathcal{F}}

\def\Tc{\mathcal{T}}

\newcommand{\tia}[1]{}

\newcommand{\bea}{\begin{eqnarray}}
\newcommand{\eea}{\end{eqnarray}}
\newcommand{\beas}{\begin{eqnarray*}}
\newcommand{\eeas}{\end{eqnarray*}}
\newcommand{\bal}{\begin{aligned}}
\newcommand{\eal}{\end{aligned}}

\def\({\left(}
\def\){\right)}

\newcommand{\pd}{\partial}

\newcommand{\const}{\mathrm{const}}

\makeatother

\begin{document}

\title{Effective models of inflation from a non-local framework}

\author{Alexey S. Koshelev}
\email{alexey@ubi.pt}

\selectlanguage{british}%

\address{Departamento de F\'{i}sica, Centro de Matem\'atica e Aplica\c{c}\~oes
(CMA-UBI), Universidade da Beira Interior, 6200 Covilh\~a, Portugal}

\address{Theoretische Natuurkunde, Vrije Universiteit Brussel and The International
Solvay Institutes, Pleinlaan 2, B-1050 Brussels, Belgium}

\address{Steklov Mathematical Institute of RAS, Gubkina str. 8, 119991 Moscow,
Russia }
\selectlanguage{english}%

\author{K. Sravan Kumar}
\email{sravan@ubi.pt}

\selectlanguage{british}%

\address{Departamento de F\'{i}sica, Centro de Matem\'atica e Aplica\c{c}\~oes
(CMA-UBI), Universidade da Beira Interior, 6200 Covilh\~a, Portugal}
\selectlanguage{english}%

\author{Paulo Vargas Moniz}
\email{pmoniz@ubi.pt}

\selectlanguage{british}%

\address{Departamento de F\'{i}sica, Centro de Matem\'atica e Aplica\c{c}\~oes
(CMA-UBI), Universidade da Beira Interior, 6200 Covilh\~a, Portugal}
\selectlanguage{english}%
\begin{abstract}
	{
The dilaton is a possible inflaton candidate following recent CMB data allowing a non-minimal coupling to the Ricci curvature scalar in the early Universe. In this paper, we introduce an approach that has seldom been used in the literature, namely dilaton inflation with non-local features. More concretely, employing non-local features expressed in J. High Energy Phys.~04 (2007) 029,
we study quadratic variations around a de Sitter geometry of an effective action with a non-local dilaton.
The non-locality
refers to an infinite derivative kinetic term involving the operator
$\mathcal{F}\left(\Box\right)$. Algebraic roots of the characteristic equation $\Fc(z)=0$ play a crucial role in determining the properties of the theory. 
We subsequently study the cases when $\mathcal{F}\left(\Box\right)$ 
has one real root and one complex root, from which we retrieve two concrete
effective models of inflation. In the first case we retrieve a class of single field inflations with universal
prediction of $n_{s}\sim0.967$ with any value of the tensor to scalar ratio $r<0.1$ intrinsically controlled by the root of the characteristic equation. The second case involves a new class
of two field conformally invariant models with a peculiar quadratic
cross-product of scalar fields. In this latter case, we obtain Starobinsky like inflation through a spontaneously
broken conformal invariance. Furthermore, an uplifted minimum of the potential, which accounts for the vacuum
energy after inflation is produced naturally through this mechanism intrinsically within our approach.}
\end{abstract}
\maketitle
\maketitle 

\section{Introduction}

\indent

Primordial inflation is a compelling paradigm for describing the early
Universe. This is manifest through convincing observational data \cite{Ade:2015lrj}.
The end of inflation is characterized by primordial perturbations
which are eventually responsible for the structure formation in the
Universe. Their characteristics, namely, spectral tilts and the ratio
of tensor to scalar power spectra $r$, have been recently measured:
$r$ has a well-established upper limit\footnote{$r<0.07$ considering the recent results of BICEP2/Keck Array \cite{Array:2015xqh}.}
$r<0.1$ at $95\%$ confidence level from Planck 2015 \cite{Ade:2015lrj,Ade:2015tva},
whereas the scalar tilt is most precisely measured as $n_{s}=0.968\pm0.006$
at $95\%$ confidence level. The CMB power spectra are so far found
to be very much adiabatic, scale invariant and Gaussian \cite{Ade:2015ava,Ade:2015lrj},
supporting thereby $f(R)$ or single field inflation models. Among
a broad class of models, the Starobinsky model based on the $R+R^{2}$
gravity modification and the Higgs inflation \cite{Starobinsky:1980te,Starobinsky:1983zz,Bezrukov:2007ep}
occupy a privileged position, with practically equal predictions in
the $\left(n_{s},\,r\right)$ plane 
$
n_{s}=1-{2}/{N}$%
,
 $r={12}/{N^{2}}$,
where $N$ is the number of $e$-foldings before the end of inflation.
For the expected value $N\approx60$ the above predictions match very
well the current observational values and constraints. However, the
physical nature of the inflaton and the corresponding mechanism driving
the early universe accelerated expansion are still an open issue \cite{Martin:2013tda,Martin:2015dha}.

It can be added that according to the present observations,
the Hubble parameter during inflation can be as large as $10^{15}$
Gev, suggesting the scale of inflation to be of the order of $M_{I}>10^{15}$
Gev. These energy scales are acceptable in supergravity (SUGRA) and
string theory, hence argued to play a crucial role \cite{Cicoli:2010yj}.
Therefore, during the last years there have been many attempts to
embed the inflationary picture into low energy effective theories
derived from such fundamental approaches \cite{Linde:2014nna,Silverstein:2013wua,Silverstein:2015mll,Burgess:2013sla,Baumann:2014nda}.
Furthermore, the observational data provided a special
stimulus to studies of inflation in SUGRA and string theory. More precisely, flat
potentials of the following form 
\begin{equation}
V\sim\left(1-e^{-\sqrt{2/3B}\varphi}\right)^{2n}\,.\label{flatpot}
\end{equation}
became successful candidates for the description of inflation and
appeared in various scenarios \cite{Martin:2013tda,Linde:2014nna,Martin:2015dha}.
The parameter $B$ in the above potential can lead to
any value of $r<0.1$ with a universal value for $n_s$ as it is in the $R^2$ model, namely
\begin{equation}
n_s=1-\frac 2N\, ,\quad r=\frac{12B}{N^{2}}\,.\label{attractorpred}
\end{equation}
Such predictions are so far shown to occur in the low energy
effective models of string theory/SUGRA and modified gravity \cite{Ellis:2013nxa,Ellis:2015xna,Kallosh:2013xya,Kallosh:2013yoa,Nastase:2015pua,Diamandis:2015xra,Ozkan:2015iva}. {In addition, several other models inspired from string theory were also successful in confronting Planck data  \cite{Blumenhagen:2015qda,Marchesano:2014mla,Westphal:2014ana,Escamilla-Rivera:2015ova,Kumar:2014oka}.}

Following the current observational status of inflation \cite{Martin:2013tda,Martin:2015dha}, a non-minimally coupled scalar was established as a suitable candidate. In this regard, it is possible consider a closed string dilaton in string theory as a candidate of interest. Embracing string theory as a key player in cosmological inflation,
we take an inspiration from string field theory (SFT) \cite{Ohmori:2001am,Arefeva:2001ps} {where non-locality can naturally emerge in the action}. Previous attempts considering inflation with non-locality features was done with $p-$adic strings \cite{Barnaby:2006hi,Biswas:2012bp}. Moreover, configurations of non-locality lead to effective field theories with one or more scalar degrees of freedom \cite{Koshelev:2007fi}. 

In the context of the previous paragraphs, we will argue that interesting inflationary scenarios can be produced with non-local features of the dilaton. 
More precisely, the non-local nature of the
dilaton is characterized by the function $\mathcal{F}\left(\Box\right)=\overset{\infty}{\underset{n=0}{\sum}}f_{n}\Box^{n}$,
where $\Box$ is the d'Alembertian.

 Depending on the number of roots
of the characteristic equation $\mathcal{F}\left(z\right)=0$, following the studies of \cite{Koshelev:2007fi,Koshelev:2009ty,Koshelev:2010bf},
we can write effective actions that are equivalent up to the quadratic
perturbations. More specifically, if $\mathcal{F}\left(\Box\right)$
has only one real root at $z_{1}$, the corresponding effective action
contains just one propagating scalar where the kinetic term contains the
parameter $\mathcal{F}^{\prime}\left(z_{1}\right)$ and any higher
derivatives can be neglected assuming the field slow-rolls on a sufficiently
flat potential. As a consequence, we can write a successful (albeit trivial) single field
inflation with controlled slow-roll dynamics through the parameter
$\mathcal{F}^{\prime}\left(z_{1}\right)$, which leads to the prediction
of $r$ in (\ref{attractorpred}). Far much more interesting if $\mathcal{F}\left(\Box\right)$
has a complex root the corresponding effective action contains instead two
real scalar fields, which we will show to bear conformal invariance. In this
case, the two scalar fields share an opposite sign of kinetic terms.
From a spontaneous breaking of conformal symmetry, we gauge fix one
of the scalar field and obtain a Starobinsky like inflation, accompanied
with a non-trivial uplifting of the inflaton potential towards a non-zero minimum.

{This paper is organized as follows. In Sec.~\ref{EFTs}, we start with a quadratic action with dilatonic perturbations around de Sitter (dS). 
We prescribe subsequently a method to write an effective
action bearing non-locality.  In Sec.~\ref{Cin} we study in detail
two particular effective actions which leads to interesting inflationary
scenarios. In Sec.~\ref{concdisc} we summarize and discuss our inflationary scenarios. 
We refer to the Appendix \ref{AppSFT} for additional notes on SFT and tachyon condensation (TC). }
{Appendix.~\ref{SFT-newA} suggest a framework concerning non-local dilaton within string theory.}

Through out the paper, we set the metric signature $(-,+,+,+)$, small
Greek letters are the fully covariant indexes and and the units $\hbar=1,\,c=1,\,M_{P}^{2}=\frac{1}{8\pi G}$.

\section{Effective fields from a non-local framework}

\label{EFTs}

{The attractor models of inflation leading to the predictions in (\ref{attractorpred}) essentially have a Starobinsky like potential\footnote{corresponding to canonical scalar field} with a parameter $B$. The realization of a Starobinsky like potential can be achieved via employing a non-minimally coupled scalar or conformal models \cite{Kallosh:2013pby,Kallosh:2013yoa}. The parameter $B$ here\footnote{i.e., so far shown to be obtained in SUGRA/string theory settings \cite{Ellis:2013nxa,Ellis:2015xna,Kallosh:2013xya,Kallosh:2013yoa,Nastase:2015pua,Diamandis:2015xra,Ozkan:2015iva}} mainly defines the coefficient of the inflaton kinetic term. In string (field) theory the kinetic term of a scalar naturally comes with analytic infinite derivative function (non-locality). Assuming there exists a non-local dilaton\footnote{That is naturally coupled to Ricci scalar} in a 4D effective version of string theory\footnote{In Appendix.~\ref{SFT-newA} we suggest a mechanism for obtaining non-local dilaton based on string (field) theory inspired set up. We defer further development of our Appendix in a subsequent study \cite{Koshelevetal}.}, we can realize the parameter $B$ by analyzing the linearization of the theory around a dS in a local limit\footnote{It is natural to expect infinite derivatives are unimportant at the inflationary energy scales}.} 

{The second order action (scalar part) of an effective theory of non-local dilaton around dS should look like}
{
\begin{equation}
\delta^{(2)}S=\frac{1}{2}\int d^{4}x\sqrt{-g}\varphi{\Fc(\Box)}\varphi\,,\label{d2sfteffFbox}
\end{equation}}

{where}
{
\begin{equation}
{\Fc}(\Box)={M_{P}^{2}}\left(2\Box+3R_{0}\right)+\tilde{\Fc}\left(\Box\right)
\end{equation}
where $\varphi$ is the perturbation of the dilatonic field ($\phi$), $R_0$ is the scalar curvature around dS, $M_p$ is the reduced Planck mass and $\tilde{\Fc}(\Box)=\sum_{n=1}^{\infty}f_n\Box^n$. Although during slow-roll inflation infinite derivatives are not so relevant, it is pertinent to identify the coefficients $f_n$.}

{To generate inflation we must have an appropriate
potential in our set-up. 
 At present, the state of the art of the knowledge in string (field) theory lacks established methods to do so. In the course
of this paper, we will employ potentials phenomenologically which we assume do not
violate general principles of string theoretical  construction (c.f. the Appendix \ref{AppSFT}, \ref{SFT-newA}
for more discussions on this issue).

Considering therefore (\ref{d2sfteffFbox}) for a general operator function
$\Fc(\Box)$ we cannot convey inflationary physics straightforwardly.
In general, $\Fc(\Box)$ being considered as an algebraic function
may have many roots. That is, equation 
\begin{equation}
\Fc(z)=0\label{characteristic}
\end{equation}
can have more than one solution. We name it a characteristic equation.
Because of that, the propagator for the field $\varphi$ will have
more than one pole. As such, it is equivalent to multiple degrees
of freedom. Let us therefore write a local realization of (\ref{d2sfteffFbox}).
Originally, this was done in \cite{Koshelev:2007fi} and then formalized
in \cite{Aref'eva:2007mf,Koshelev:2009ty,Koshelev:2010bf}. We use
the Weierstrass factorization \cite{Koshelev:2007fi} which prescribes
that any entire function (we recall that SFT ensures that operators
$\Fc(\Box)$ are analytic functions and in all existing computations they appear to be entire functions) can be written as 
\begin{equation}
\Fc(z)=e^{\gamma(z)}\prod_{j}\left(z-z_{j}\right){}^{m_{j}}\,,\label{weierstrass}
\end{equation}
where $z_{j}$ are roots of the characteristic equation and $m_{j}$
are their respective multiplicities. We further assume hereafter that all
$m_{j}=1$ for simplicity. $\gamma(z)$ is an entire function and
as such its exponent has no roots on the whole complex plane. It was
shown in \cite{Koshelev:2007fi} that for a quadratic Lagrangian of
the type (\ref{d2sfteffFbox}), a local equivalent quadratic Lagrangian
can be constructed as 
\begin{equation}
\delta^{2}S_{local}=\frac{1}{2}\int d^{4}x\sqrt{-g}\sum_{j}\Fc'\left(z_{j}\right)\varphi_{j}\left(\Box-z_{j}\right)\varphi_{j}\,,\label{locallocal}
\end{equation}
where prime means derivative with respect to the argument $z$ with
the further evaluation at the point $z_{j}$. It is said to be equivalent because of the fact that solution for $\varphi$, which can be obtained
from equations of motion following from (\ref{d2sfteffFbox}), is connected
to solutions for $\varphi_{j}$ simply as
\begin{equation}
\varphi=\sum\varphi_{j}\,.\label{solsol}
\end{equation}

Roots $z_{j}$ become therefore the most crucial elements and several comments are in order here: 
\begin{itemize}
\item Note that roots $z_{j}$ can be complex in general. One real root
$z_{1}$ is the simplest situation (c.f. Sec.~\ref{Bsing}). In this case, we have just a Lagrangian
for a massive scalar. It is acceptable if $\Fc'(z_{1})>0$ in order
to evade a ghost in the spectrum.
\item More than one real root apparently seems not to be a promising scenario.
Since the function $\Fc(z)$ is analytic (and therefore continuous),
neighbouring real roots will be accompanied with $\Fc'\left(z_{j}\right)$
of opposite signs. In other words, one root is normal and the next
to it is a ghost. We study an effective model corresponding to this case in Sec.~\ref{Cin}.
\end{itemize}

\subsection{Effective model of single field inflation}

\label{Bsing}

If $\mathcal{F}\left(z\right)$ has one real root, then (\ref{locallocal})
contains a single scalar degrees of freedom 

\begin{equation}
\delta^{2}S_{local}=\frac{1}{2}\int d^{4}x\sqrt{-g}\Fc'\left(z_{1}\right)\varphi\left(\Box-z_{1}\right)\varphi\label{eff-single-field}
\end{equation}
The effective action which is perturbatively equivalent up to quadratic
order to (\ref{eff-single-field}) around dS background, looks
like (taking $M_{P}=1$)

\begin{equation}
S_{1}=\int d^{4}x\sqrt{-g}\left[\frac{1}{2}\tilde{\Phi}^{2}R-\frac{A}{2}\pd\tilde{\Phi}^{2}-V(\tilde{\Phi})\right]\,,\label{example1}
\end{equation}
where $\tilde{\Phi}$ is an effective dilatonic field and the respective correspondence is 

\begin{equation}
\begin{split}\Fc'(z_{1}) & =6+A\\
\Fc'(z_{1})z_{1} & =3R_{0}-V''\left(\tilde{\Phi}_{0}\right)\,.
\end{split}
\label{identification}
\end{equation}
Here $ R_{0} $ is scalar curvature of the dS vacuum solution for a constant $\tilde{\Phi}$. 
Assuming the generalized structure of from the proposed action (\ref{action_model_new}),
the potential $V(\tilde{\Phi})$ can be taken to be arbitrary. 
If we consider a potential $V_{J}\left(\tilde{\Phi}\right)=V_{0}\left(-\tilde{\Phi}^{2}+\tilde{\Phi}^{4}\right)^{2}$ which looks in the Einstein frame as 
\begin{equation}
V_{E}=\tilde{V}_{0}\left(1-e^{-\sqrt{\frac{2}
		{3\left[\mathcal{F}^{\prime}\left(z_{1}\right)/6\right]}}
	\tilde{\phi}}\right)^{2}\,,\label{potEmodel}
\end{equation}
where $\tilde{\phi}$ is canonically normalized field by definining $\tilde{\Phi}=e^{-\sqrt{\frac{1}{A+6}}\tilde{\phi}}$. 
The inflationary predictions corresponding to
the potential in (\ref{potEmodel}) are well known \cite{Ellis:2013nxa,Kallosh:2013yoa,Kehagias:2013mya,Carrasco:2015pla} and in particular we retrieve 
\[
n_{s}=1-\frac{2}{N}\quad,\quad r=\frac{2\mathcal{F}^{\prime}\left(z_{1}\right)}{N^{2}}\,,
\]
where we consider  $N=60$ number of $e$-foldings. 
We therefore conclude that provided the non-local operator $\Fc(\Box)$
contains one real root, it gives a successful inflation with a universal
prediction of $n_{s}=0.967$ and the tensor
to scalar ratio $r<0.1$. The value of $r$ can be varied to any value
by varying the non-local parameter $\mathcal{F}^{\prime}\left(z_{1}\right)$.

\subsection{Effective model of conformal inflation}

\label{Cin}

If $\mathcal{F}\left(z\right)$ has a complex root then we should
write (\ref{locallocal}) for a scalar field and also for its
complex conjugate. So considering such a pair of complex conjugate
roots, we have 
\begin{equation}
\delta^{2}S_{local}=\frac{1}{2}\int d^{4}x\sqrt{-g}\left[\Fc'\left(z_{1}\right)\varphi_{1}\left(\Box-z_{1}\right)\varphi_{1}+\Fc'\left(\bar{z}_{1}\right)\bar{\varphi}_{1}\left(\Box-\bar{z}_{1}\right)\bar{\varphi}_{1}\right]\,,\label{pairpair}
\end{equation}
where a bar over represents the complex conjugates. To maintain the
connection with (\ref{d2sfteffFbox}) we should
consider complex conjugate solutions to equations of motion, such
that $\varphi=\varphi_{1}+\bar{\varphi}_{1}$ is real. The important
feature is that the quadratic form of fields is already diagonal.
Introducing $\varphi_{1}=\chi+i\sigma$, $z_{1}=\alpha+i\beta$, $\Fc'(z_{1})=c+is$
we can rewrite (\ref{pairpair}) in terms of real components
as 
\begin{equation}
\delta^{2}S_{local}=\int d^{4}x\sqrt{-g}\left[\chi(c\Box-c\alpha+s\beta)\chi-\sigma(c\Box-c\alpha+s\beta)\sigma-2\chi(s\Box-s\alpha-c\beta)\sigma\right]\label{pairpairreal}
\end{equation}
The above expression is inevitably non-diagonal and features a cross-product
of real fields $\sim\chi\sigma$. In this formulation, note that
the two fields $\chi,\,\sigma$ share an opposite sign concerning their kinetic terms
\cite{Galli:2010qx}.  

Let us now show that the following effective action
with two fields with conformal invariance, can be perturbatively equivalent
up to quadratic order to (\ref{pairpairreal}) around dS background
\begin{equation}
\begin{split}S_{2}=\int d^{4}x\sqrt{-g} & \left[\frac{M_{P}^{2}}{2}[\tilde{\alpha}\tilde{\Phi}_{1}^{2}-\tilde{\alpha}\tilde{\Phi}_{2}^{2}{-}2\tilde{\beta}\tilde{\Phi}_{1}\tilde{\Phi}_{2}]f\left(\frac{\tilde{\Phi}_{2}}{\tilde{\Phi}_{1}}\right)R\right.\\
 & +\left.\frac{A}{2}[\tilde{\alpha}\pd\tilde{\Phi}_{1}^{2}-\tilde{\alpha}\pd\tilde{\Phi}_{2}^{2}-2\tilde{\beta}\pd_{\mu}\tilde{\Phi}_{1}\pd^{\mu}\tilde{\Phi}_{2}]f\left(\frac{\tilde{\Phi}_{2}}{\tilde{\Phi}_{1}}\right)-V\left(\tilde{\Phi}_{1},\tilde{\Phi}_{2}\right)\right]\,.
\end{split}
\label{example2}
\end{equation}
where $\tilde{\Phi}_{1},\,\tilde{\Phi}_{2}$ are effective dilatonic fields. 

We can write the quadratic Lagrangian for the spin-0 part, which contains
2 components $\tilde{\chi}{=\delta\tilde{\Phi}_{1}}$ and $\tilde{\sigma}{={\delta\tilde{\Phi_{2}}}}$
(i.e. again the spin-0 metric perturbation is excluded by equations
of motion), as 
\begin{equation}
\begin{split}\delta^{2}S_{2}=\frac{1}{2}\int d^{4}x\sqrt{-g} & \left[\tilde{\chi}\Delta_{\tilde{\chi}}\tilde{\chi}+\tilde{\sigma}\Delta_{\tilde{\sigma}}\tilde{\sigma}+\tilde{\chi}\Delta_{\tilde{\chi}\tilde{\sigma}}\tilde{\sigma}\right]\end{split}
\,,\label{s2eff}
\end{equation}
where 
\begin{eqnarray*}
\Delta_{\tilde{\chi}} & = & \frac{M_{P}^{2}}{2}\left(\frac{(\pd_{\tilde{\Phi}_{1}}I_{0})^{2}}{I_{0}}(3\Box+R_{0})+\frac{\pd^{2}I_{0}}{\pd\tilde{\Phi}_{1}^{2}}R_{0}\right)-A\tilde{\alpha}f_{0}\Box-\frac{\pd^{2}V_{0}}{\pd\tilde{\Phi}_{1}^{2}}\,,\\
\Delta_{\tilde{\sigma}} & = & \frac{M_{P}^{2}}{2}\left(\frac{(\pd_{\tilde{\Phi}_{2}}I_{0})^{2}}{I_{0}}(3\Box+R_{0})+\frac{\pd^{2}I_{0}}{\pd\tilde{\Phi}_{2}^{2}}R_{0}\right)+A\tilde{\alpha}f_{0}\Box-\frac{\pd^{2}V_{0}}{\pd\tilde{\Phi}_{2}^{2}}\,,\\
\Delta_{\tilde{\chi}\tilde{\sigma}} & = & \frac{M_{P}^{2}}{2}\left(\frac{\pd_{\tilde{\Phi}_{1}}I_{0}\pd_{\tilde{\Phi}_{2}}I_{0}}{I_{0}}(3\Box+R_{0})+\frac{\pd^{2}I_{0}}{\pd\tilde{\Phi}_{1}\pd\tilde{\Phi}_{2}}R_{0}\right)-A\tilde{\beta}f_{0}\Box-\frac{\pd^{2}V_{0}}{\pd\tilde{\Phi}_{1}\pd\tilde{\Phi}_{2}}\,,
\end{eqnarray*}
where $R_{0}$ is the scalar curvature of dS vacuum for constant dilatonic fields $\tilde{\Phi}_{1}=\tilde{\Phi}_{1,0},\,\tilde{\Phi}_{2}=\tilde{\Phi}_{2,0}$. Here we define $I(\tilde{\Phi}_{1},\tilde{\Phi}_{2})=\left[\tilde{\alpha}\tilde{\Phi}_{1}^{2}-\tilde{\alpha}\tilde{\Phi}_{2}^{2}{-}2\tilde{\beta}\tilde{\Phi}_{1}\tilde{\Phi}_{2}\right]f\left({\tilde{\Phi}_{2}}/{\tilde{\Phi}_{1}}\right)$
and $I_{0}\equiv I(\tilde{\Phi}_{1,0},\tilde{\Phi}_{2,0})$, 
$\pd_{\tilde{\Phi}_{1}}I_{0}\equiv\pd I(\tilde{\Phi}_{1},\tilde{\Phi}_{2})/\pd\tilde{\Phi}_{1}$
are the quantities evaluated at the values of fields at dS vacuum and so on for analogous terms. 

We make use of (\ref{pairpairreal}), in the case of two
complex conjugate roots with the Lagrangian written in real fields.
Hence, we juxtapose (\ref{pairpairreal}) and (\ref{s2eff}).
The motivation for doing this is to establish a more fundamental correspondence\footnote{This is, however much more
	involved than in Appendix.~\ref{Bsing} with a single field.}
for the effective model (\ref{example2}). Essentially,
the most important is to establish $\Delta_{\tilde{\chi}}=-\Delta_{\tilde{\sigma}}$.
In this manner, we can neglect the second derivatives of the potential
$V$. However, we must satisfy a number of constraints, namely, all
parameters and vacuum fields values must be real and $I_{0}$ strictly
positive. And we want to have $\tilde{\beta}\neq0$, which we will satisfy in the following. The greatly simplifying point is that
we must require such an adjustment of coefficients of $\Delta$-s
only in a single point $(\tilde{\Phi}_{1}=\tilde{\Phi}_{1,0}$, $\tilde{\Phi}_{2}=\tilde{\Phi}_{2,0})$.
On top of this we emphasize again that we aim at retrieving a
nearly dS phase, not an exact one. These requirements are generically
satisfied altogether with the presence of a function $f\left(\frac{\tilde{\Phi}_{2}}{\tilde{\Phi}_{1}}\right)$
(apart from special situations which we discuss shortly). It is important
that being a function of the ratio of fields it cannot spoil a possible
conformal invariance.

Our argument and the construction is to establish
an effective setting which can emulate (\ref{pairpairreal}). We claim
that we have such an effective framework as long we can match quadratic
actions for scalar modes around a dS background. We can thus establish
a correspondence between (\ref{pairpairreal}) and (\ref{s2eff})
by means of the following: 
\begin{itemize}
\item During inflationary expansion we can assume that the scalar fields
vary slowly and the kinetic terms can be neglected. We are thus
mainly interested in whether $\Delta_{\tilde{\chi}}=-\Delta_{\tilde{\sigma}}$
for the terms proportional to $R_{0}$. To have this we should require
\begin{equation}
\frac{(\pd_{\tilde{\Phi}_{1}}I_{0})^{2}}{I_{0}}+\frac{\pd^{2}I_{0}}{\pd\tilde{\Phi}_{1}^{2}}+\frac{(\pd_{\tilde{\Phi}_{2}}I_{0})^{2}}{I_{0}}+\frac{\pd^{2}I_{0}}{\pd\tilde{\Phi}_{2}^{2}}\approx0\label{condr0}
\end{equation}
\item We can check that even in the very simple case of $\tilde{\beta}=0$,
a non-constant function $f$ is required to satisfy the above relation.
A simple choice like 
\begin{equation}
f=1+f_{1}\tilde{\Phi}_{2}/\tilde{\Phi}_{1}\,,\label{condr0f}
\end{equation}
with just one free parameter $f_{1}$ is sufficient. Otherwise, for
$f=\const$ a condition $\tilde{\Phi}_{1,0}=\pm i\tilde{\Phi}_{2,0}$
arises from (\ref{condr0}). Therefore to build such an effective
model the function $f\left(\frac{\tilde{\Phi}_{2}}{\Phi_{1}}\right)$
is very useful and important. The cross-product of fields may arise
for $\tilde{\beta}=0$ but a quite involved non-polynomial function
$f$ is required. 
\item For a non-trivial $\tilde{\beta}$ the same function $f$ as above
in (\ref{condr0f}) is enough to arrange the condition (\ref{condr0}).
Moreover $\tilde{\beta}\neq0$ generates a cross-product of fields. 
\item In complete analogy we can consider the coefficients of the kinetic
terms. We have to require a non-constant function $f$. We note that
having opposite coefficients in front of d'Alembertian operators for
different fields essentially means that one of these fields is a ghost. 
\end{itemize}
Recalling expressions (\ref{locallocal}) and (\ref{pairpairreal}),
we see that the presence of a cross-product is a special feature related
to a complex root of the function $\Fc(z)$ (which defines the non-local
operator $\Fc(\Box)$). This means that the parameter $\beta$ found
in (\ref{pairpairreal}) is essentially non-zero (notice that there
is no direct simple relation between $\tilde{\beta}$ and $\beta$).
In the limiting case of $\beta\to0$, we should see the cross-product
disappearing and this corresponds to $\tilde{\beta}\to0$ in the effective
model (\ref{example2}). Another way to recognize the effective model
(\ref{example2}) without a cross-product of fields is to consider
directly (\ref{locallocal}) with two specially tuned real roots.
This means that these roots are related as $z_{2}=-z_{1}$ and moreover
$\Fc'(z_{2})=-\Fc'(z_{1})$.

To resolve the issue of a ghost in the spectrum requires an extra
symmetry in order to gauge the ghost away. The most natural candidate
is the conformal symmetry used in the building of similar models in
\cite{Kallosh:2013xya,Kallosh:2013pby,Kallosh:2013yoa}. The conformal
invariance is restored in (\ref{example2}) if we assume $A=6$. Our
model without a cross-product resembles the conformal models studied
in \cite{Kallosh:2013hoa,Kallosh:2013daa}. We stress that the cross-product
appeared for the first time in the cosmological models and we have
here provided an imperative explanation through the non-local dilaton.

Assuming $f\left(\frac{\tilde{\Phi}_{2}}{\tilde{\Phi}_{1}}\right)\approx\mathrm{constant}$
during inflation, then (\ref{example2}) can be written as
\begin{equation}
\begin{split}S_{2}=\int d^{4}x\sqrt{-g} & \left[\left(\tilde{\alpha}\tilde{\Phi}_{1}^{2}-\tilde{\alpha}\tilde{\Phi}_{2}^{2}{-}2\tilde{\beta}\tilde{\Phi}_{1}\tilde{\Phi}_{2}\right)\frac{R}{12}\right.\\
 & +\left.\frac{\tilde{\alpha}}{2}\pd\tilde{\Phi}_{1}^{2}-\frac{\tilde{\alpha}}{2}\pd\tilde{\Phi}_{2}^{2}-\tilde{\beta}\pd_{\mu}\tilde{\Phi}_{1}\pd^{\mu}\tilde{\Phi}_{2}-V_{J}\left(\tilde{\Phi}_{1},\tilde{\Phi}_{2}\right)\right]\,,
\end{split}
\label{Cinf}
\end{equation}
where we have set $M_{P}=1$ for simplicity and use the subscript
$J$ for the Jordan frame as before. Since the field $\tilde{\Phi}_{1}$
has a wrong sign kinetic term (assuming $\tilde{\alpha}>0$), we can
eliminate it by the choice of conformal gauge $\tilde{\Phi}_{1}=\sqrt{6}$
which spontaneously breaks the conformal invariance. To obtain a consistent
inflation within this model we consider the following potential 
\begin{equation}
V_{J}\left(\tilde{\Phi}_{1},\,\tilde{\Phi}_{2}\right)=\frac{\lambda}{4}\left(\gamma_{1}\tilde{\Phi}_{2}^{2}+\gamma_{2}\tilde{\Phi}_{1}\tilde{\Phi}_{2}+\gamma_{3}\tilde{\Phi}_{1}^{2}\right)\left(\tilde{\Phi}_{2}-\tilde{\Phi}_{1}\right)^{2}\,,\label{inflatonpotCI}
\end{equation}
where $\gamma_{1}$, $\gamma_{2}$, $\gamma_{3}$ are arbitrary constant
parameters. The potential (\ref{inflatonpotCI}) is motivated from
\cite{Kallosh:2013xya}, which we generalize here to our conformal
model with a term containing the cross-product of fields. The importance
of this generalization will be explained in what follows. Note that
if $\tilde{\beta}=\gamma_{2}=\gamma_{3}=0$ , the model reduces to
the conformal model without a cross-product of fields studied in \cite{Kallosh:2013xya}.

Rescaling the fields as $\tilde{\Phi}_{1}\to\frac{\tilde{\Phi}_{1}}{\sqrt{\tilde{\alpha}}}$
and $\tilde{\Phi}_{2}\to\frac{\tilde{\Phi}_{2}}{\sqrt{\tilde{\alpha}}}$
in action (\ref{Cinf}) and using the gauge $\tilde{\Phi}_{1}=\sqrt{6}$
we yield

\begin{equation}
\begin{split}S_{2}=\int d^{4}x\sqrt{-g} & \left[\frac{R}{2}\left(1-\frac{\tilde{\Phi}_{2}^{2}}{6}-\frac{2\tilde{\beta}}{\sqrt{6}\tilde{\alpha}}\tilde{\Phi}_{2}\right)\right.\\
 & \left.-\frac{1}{2}\partial_{\mu}\tilde{\Phi}_{2}\partial^{\mu}\tilde{\Phi}_{2}-\frac{\lambda}{4\tilde{\alpha}^{2}}\left(\gamma_{1}\tilde{\Phi}_{2}^{2}+\gamma_{2}\tilde{\Phi}_{1}\tilde{\Phi}_{2}+\gamma_{3}\tilde{\Phi}_{1}^{2}\right)\left(\tilde{\Phi}_{2}-\sqrt{6}\right)^{2}\right]\,.
\end{split}
\label{sft3}
\end{equation}



Performing the conformal transformation $g_{\mu\nu}\to\left[1+\frac{\tilde{\beta}^{2}}{\tilde{\alpha}^{2}}-\frac{1}{6}\left(\tilde{\Phi}_{2}+\frac{\tilde{\beta}}{\tilde{\alpha}}\sqrt{6}\right)^{2}\right]^{-1}g_{\mu\nu}$
and shifting the field $\tilde{\Phi}_{2}\to\tilde{\Phi}_{2}+\frac{\tilde{\beta}}{\tilde{\alpha}}\sqrt{6}$,
we arrive to the Einstein frame action 
\begin{equation}
S_{2E}=\int d^{4}x\,\sqrt{-g_{E}}\left[\frac{R_{E}}{2}-\frac{\omega}{2\left(\omega-\frac{\tilde{\Phi}_{2}^{2}}{6}\right)^{2}}\partial_{\mu}\tilde{\Phi}_{2}\partial^{\mu}\tilde{\Phi}_{2}-V_{E}\left(\tilde{\Phi}_{2}\right)\right]\,,\label{sft5}
\end{equation}
where $\omega=1+\frac{\tilde{\beta}^{2}}{\tilde{\alpha}^{2}}$ and
\begin{equation}
V_{E}\left(\tilde{\Phi}_{2}\right)=\frac{9\lambda}{\tilde{\alpha}^{2}}\frac{\left[\gamma_{1}\tilde{\Phi}_{2}^{2}+\left(\gamma_{2}-2\gamma_{1}\frac{\tilde{\beta}}{\tilde{\alpha}}\right)\sqrt{6}\tilde{\Phi}_{2}+6\left(\gamma_{1}\frac{\tilde{\beta}^{2}}{\alpha^{2}}-\gamma_{2}\frac{\tilde{\beta}}{\tilde{\alpha}}+\gamma_{3}\right)\right]\left(\tilde{\Phi}_{2}-\sqrt{6}\frac{\tilde{\beta}}{\tilde{\alpha}}-\sqrt{6}\right)^{2}}{\left(6\omega-\tilde{\Phi}_{2}^{2}\right)^{2}}\,.\label{finalEpot}
\end{equation}
If $\gamma_{i}$ are chosen such that $\gamma_{2}=2\gamma_{1}\frac{\tilde{\beta}}{\tilde{\alpha}}$
and $\gamma_{1}\frac{\tilde{\beta}^{2}}{\alpha^{2}}-\gamma_{2}\frac{\tilde{\beta}}{\tilde{\alpha}}+\gamma_{3}\gtrsim0$,
we can obtain inflation with an uplifting of the potential at the
minimum.

Being more concrete, let us consider a simple case with $\gamma_{1}=1$ ,
$\gamma_{2}=2\frac{\tilde{\beta}}{\tilde{\alpha}}$ and $\gamma_{3}=2\frac{\tilde{\beta}^{2}}{\tilde{\alpha}^{2}}$
, for which (\ref{finalEpot}) reduces to the following form interms of canonically normalized field $\tilde{\Phi}_{2}=\sqrt{6\omega}\tanh\left(\frac{\tilde{\phi}}{\sqrt{6}}\right)$
\begin{equation}
\begin{aligned}V_{E}\left(\tilde{\phi}\right)= & \mu^{2}\left[\sinh^{2}\left(\frac{\tilde{\phi}}{\sqrt{6}}\right)+\frac{\tilde{\beta}^{2}}{
\left(\tilde{\alpha}^{2}+\tilde{\beta}^{2}\right)}\cosh^{2}\left(\frac{\tilde{\phi}}{\sqrt{6}}\right)\right]\left[\cosh\left(\frac{\tilde{\phi}}{\sqrt{6}}\right)
-\frac{1}{1+\frac{\tilde{\beta}}{\tilde{\alpha}}}\sqrt{1+\frac{\tilde{\beta}}{\tilde{\alpha}^{2}}^{2}}\sinh\left(\frac{\tilde{\phi}}{\sqrt{6}}\right)\right]^{2}\\
\end{aligned}
\label{final-pot}
\end{equation}
where $\mu^{2}=\frac{9\lambda\left(\tilde{\alpha}+\tilde{\beta}\right)^{2}}{\tilde{\alpha}^{2}\left(\tilde{\alpha}^{2}+\tilde{\beta}^{2}\right)}$.
In the limit $\frac{\tilde{\beta}}{\tilde{\alpha}}\ll1$ , the first
term in (\ref{final-pot}) dominates during inflation while the second
term is negligible. The potential (\ref{final-pot}) is always positive
and in particular has a non-zero value at the minimum at $\tilde{\phi}\approx0$.
In general the shape of the potential is similar to the Starobinsky-like
models in no-scale SUGRA \cite{Ellis:2013nxa}.


Setting $\tilde{\alpha}=1$, in the limit $\tilde{\beta}\ll1$, we can approximate
the potential in (\ref{final-pot}) as
\begin{equation}
V_{E}\left(\tilde{\phi}\right)\approx\frac{\mu^{2}}{4}\left(1-e^{-\sqrt{\frac{2}{3}}\tilde{\phi}}\right)^{2}+\frac{\mu^{2}\tilde{\beta}^{2}}{4}\left(1+e^{-\sqrt{\frac{2}{3}}\tilde{\phi}}\right)^{2}\,,\label{appfinalpot}
\end{equation}
where the first term dominates when $\tilde{\phi}\gg1$ and leads
to a Starobinsky like inflation i.e., $n_{s}\sim0.967,\,r\sim0.0033$
for $N=60$ and the second term gives a non-zero vacuum energy at
the minimum of the potential\footnote{A potential of similar kind can be found in the $\alpha-$attractor
models where the inflaton potential was uplifted due to the effect
of a SUSY breaking mechanism \cite{Carrasco:2015pla}.} near $\tilde{\phi}=0$. Here $\mu\approx2\times10^{-5}$ (in Planck
units as we have set $M_{P}=1$) which can be determined from the
observed amplitude of scalar perturbations $A_{s}=2.2\times10^{-9}$
at the horizon exit \cite{Ade:2015lrj}. In particular $\tilde{\beta}\sim10^{-55}$
gives a vacuum energy that reproduces the present day cosmological
constant $\Lambda\sim10^{-120}$. Therefore, we conclude that a non-locally
induced cross-product of the fields $\tilde{\Phi}_{1}$ and $\tilde{\Phi}_{2}$
in (\ref{Cinf}) naturally uplifts the inflaton potential at the minimum
and possibly explain the present day dark energy (assuming it is $\Lambda\text{CDM}$).

\section{Conclusions and discussion}

\label{concdisc}

The dilaton is a possible inflaton candidate in the view of latest CMB data endorsing a non-minimal coupling to the Ricci curvature scalar. In this paper we investigated effective models of inflation emerged from an assumed second order action of non-local dilaton around dS. Here the non-locality enters in the dilaton kinetic term with an analytic infinite derivative function $\Fc(\Box)$. 
Using the non-local features explained in \cite{Koshelev:2007fi}, 

We analyze the cases corresponding to the roots $z_{j}$ of the characteristic equation $\Fc(z)=0$. The presence of the cross-product is a special feature related to a complex root of the function $f(z)$ (which defines the non-local operator $f(\Box)$) . Moreover, the derivatives $\Fc'(z_{j})$ play an important role.  This is seen from action (\ref{locallocal}), which
describes the evolution of scalar perturbations around a dS vacuum
within a non-local context, non-locality being a guide in this process. Its
importance is obvious as inflation is a dS like expansion and all
the observable quantities related to scalars can be obtained from
exploring the action for linear perturbations. A very important restriction
is that no ghosts must be in the spectrum. This selects two configurations
of roots.

First, there is a situation with one real root $z_{1}$ accompanied
with a correct sign of $\Fc'(z_{1})$. In this case there is one scalar
perturbative degree of freedom. Such a configuration can be obtained
from the effective model description (\ref{example1}). It is important
that coefficients in front of the Einstein-Hilbert term and the kinetic
term of a scalar field are independent. We therefore conclude that provided the non-local operator $\Fc(\Box)$
contains one real root, it gives a successful inflation with a universal
prediction of $n_{s}=0.967$ and tensor
to scalar ratio as in (\ref{attractorpred}) which can be adjusted to any
value $r<0.1$ by means of the parameter
$\mathcal{F}^{\prime}\left(z_{1}\right)$.
A future more accurate detection of parameter $r$ from CMB \cite{Creminelli:2015oda} would indicate the values of $z_{1}$
and $\Fc'(z_{1})$.

Second, there was a case with two roots. They can be complex conjugate
and then we should look at (\ref{pairpairreal}) which is written
in manifestly real components. In this scenario, we inevitably get
a quadratic cross-product of fields. Moreover, one field looks like a ghost.
However, kinetic and mass terms have exactly opposite signs. This
suggests that a conformal symmetry may help exorcizing the ghost.
Indeed, building an effective model (\ref{example2}) we have taken
the conformal symmetry into account and have shown that we indeed
can make use of it to remove the unwanted degrees of freedom. The
cross-product of fields naturally leads to an uplifting of the potential
in the reheating point. In principle one can get a similar two-field
model starting with two real roots which are related as $z_{1}=-z_{2}$
and $\Fc'(z_{1})=-\Fc'(z_{2})$. This latter case has no cross-product
of fields and falls into the considerations of \cite{Kallosh:2013hoa,Kallosh:2013daa}.
The novel feature here is that the conformally invariant models with
a quadratic cross-product of scalar fields appear for the first time
in a cosmological setup and can be naturally explained using the non-locality
of a dilaton.

More generic configurations with more than two fields may have no
reasonably simple effective model counterpart. This is because more
than one ghost would appear. In this case quite a peculiar structure
may be required in order to arrange such a configuration that it will
be possible to gauge away all the ghosts. However, although understanding
a potential power of multifield models \cite{Kallosh:2013daa}, we
surely leave this as an open question. We also have skipped a case
of multiple roots. It can be considered analogously but requires a
more complicated formula mirroring (\ref{locallocal}).

We note that models of inflation obtained in this paper can be distinguished
upon a deeper study of bi-spectrum and/or the reheating phase. This
is because in such computations full non-local operators will come
into play and these structures are unique for the presently studied
class of models.



\acknowledgments

We would like to thank Sergey Vernov for useful comments. This research
work is supported by the grant UID/MAT/00212/2013. AK is supported by Funda\c{c}\~ao para a Ci\^encia e Tecnologica (FCT) Portugal investigator project IF/01607/2015, FCT
Portugal fellowship SFRH/BPD/105212/2014 and in part by FCT Portugal
grant UID/MAT/00212/2013. SK acknowledges the support from the FCT grant
SFRH/BD/51980/2012. The authors acknowledge the COST Action CA15117
(CANTATA).



\appendix

\section{A review of SFT and Tachyon condensation}

\label{AppSFT}

In generic words SFT is an off-shell description of interacting strings
\cite{Witten:1985cc,Witten:1986qs,Zwiebach:1993cs,Ohmori:2001am,Arefeva:2001ps,Berkovits:1998bt,Berkovits:2004xh}.
It describes a string by means of a string field $\Psi$. This object
is a shorthand for encoding all the string excitations in one instance.
The corresponding action for open string field\footnote{An action for a closed SFT can be written only in a non-polynomial
form, even for the bosonic strings \cite{Saadi:1989tb,Sonoda:1989sj}.} can be written as 
\begin{equation}
S=\frac{1}{g_{o}^{2}}\left(\frac{1}{2}\int\Psi\star Q\Psi+\frac{1}{3}\int\Psi\star\Psi\star\Psi\right)\label{osft}
\end{equation}
where $\star$ and $\int$ are Witten product and integral for string
fields respectively. $Q$ is the Becchi-Rouet-Stora-Tyutin (BRST) charge. The first term clearly
corresponds to the motion of free strings while the second term represents
the interaction. The second term is the three-string vertex responsible
for the non-perturbative physics. $g_{o}$ is the open string coupling
constant, it is dimensionless.

It has been understood \cite{Sen:1998sm,Sen:1999xm,Sen:1999nx,Berkovits:2000hf,Aref'eva:2000mb}
that the tachyon of open strings is responsible for the decay of unstable
$D$-branes or $D$-brane-anti-$D$-brane pairs. The corresponding
process is the condensation of the tachyon (TC) to a non-perturbative
minimum\footnote{The TC process itself
	does not require a dynamical departure from a Minkowski background.
	This is supported by explicit papers \cite{Moeller:2002vx,Aref'eva:2003qu}
	and related studies.}. Upon the TC the unstable brane (or pair) decays. It is the
cornerstone of Sen's conjecture regarding TC that the depth of the
tachyon potential minimum is exactly the tension of an unstable brane
to which the string is attached to. The decay of a brane represents
a configuration in which open strings must not exist, because the
brane, to which they were attached, has decayed \cite{Rastelli:2000hv,Rastelli:2001jb}.
This being said, let us assume Sen's conjecture, which prescribes
the disappearance of open string excitations. The latter phenomenon
of open strings extinction can be formalized as follows in the field-theoretical
language. Given a field $\varphi$ the following quadratic Lagrangians
are non-dynamical 
\begin{equation}
L=-m^{2}\varphi^{2}\text{ or }L=\varphi e^{\gamma(\Box)}\varphi\label{nondyn}
\end{equation}
The left Lagrangian is clearly a mass term without any dynamics. In
the right Lagrangian, $\Box$ is the space-time d'Alembertian and
$\gamma$ is an entire function. Although it may look like $\Box$
produces dynamics as it is a differential operator, as long as we
require that the function in the exponent is an entire function, the
whole exponent has no eigenvalues as an operator. This means that
the inverse of such an exponent gives no poles in the propagator and
effectively we have no dynamics at all.

We further notice that the right Lagrangian in (\ref{nondyn}) is
an essentially non-local Lagrangian. It is obviously non-dynamical
on the quadratic level and as long as the field $\varphi$ is alone.
However, novel and unusual effects can be generated upon coupling
to other fields or in the non-linear physics \cite{Sen:2002nu,Moeller:2002vx,Aref'eva:2003qu,Barnaby:2006hi,Koshelev:2007fi}.

The essence of SFT is that as long as a string interaction is involved
then the non-locality of the above type emerges. Technically, we can
understand this as follows. Strings are extended objects by construction.
When a field-theoretic model describes strings, this property of an
extended object is encoded in the non-locality of interactions. SFT
straightforwardly creates vertex terms of the form 
\begin{equation}
\sim\left(e^{\alpha'\Box}\varphi_{1}\right)\left(e^{\alpha'\Box}\varphi_{2}\right)\left(e^{\alpha'\Box}\varphi_{3}\right)\label{sftvertex}
\end{equation}
Here $\alpha'$ is the string length squared (which may be different
from the inverse of the Planck mass squared).

Note
that upon lengthy computations \cite{Ohmori:2001am,Arefeva:2001ps}, the quadratic
Lagrangian of the open string tachyon $\Tc$ near the vacuum is non-dynamical
of the form 
\begin{equation}
L_{\Tc}=-\frac{{T}}{2}v(\Box,\Tc)\,.\label{tachyonnearvac}
\end{equation}
For zero momenta, i.e. when $\Box=0$ the resulting $v(0,\Tc)$ is
exactly the tachyon potential. The dependence on $\Box$ is analytic
and being linearized near the vacuum value of field $\Tc=\Tc_0+\tau$ it produces
\begin{equation}
L_{\tau}=-\frac{{T}}{2}\frac{v''(\Tc=\Tc_{0})}{2}\tau e^{\gamma(\Box)}\tau\,,\label{tachyonvac}
\end{equation}
with some entire function $\gamma(\Box)$. The coupling ${T}$ is
nothing but the tension of the unstable $D$-brane given as 
\begin{equation}
{T}=\frac{1}{2\pi^{2}g_{o}^{2}(\alpha')^{\frac{p+1}{2}}}\,,\label{tachyoncoupling}
\end{equation}
where $\alpha'$ is the string length squared, $g_o$ is the open string coupling constant and $p$ comes from
the dimensionality of the $Dp$-brane. Thus, as expected for a 3-brane,
$T$ has a dimension $[\mathrm{length}]^{-4}$ and the tachyon field
$\tau$ is dimensionless.

\section{A SFT inspired framework for non-local dilaton}

\label{SFT-newA}
 
Let us start with the well-known action of a low energy open-closd SFT coupling, obtained in the framework of the linear dilaton conformal field theory \cite{Gasperini:2007ar,Koshelev:2007fi} (see for instance \cite{Aref'eva:2008gj}).

\begin{equation}
S=\int d^{4}x\sqrt{-g}\left[\frac{M_{P}^{2}}{2}\left(\Phi^{2}R+4\pd_{\mu}\Phi\pd^{\mu}\Phi\right)-\frac{{T}}{2}\Phi[v(\Box,\Tc)+1]\right].\label{action_model}
\end{equation}
where we have redefined the dilaton field as $\Phi=e^{-\phi}$. In the above system, tachyon is assumed to be near the potential minimum and its dynamics can be neglected (see Appendix \ref{AppSFT}). 
A careful but quick analysis immediately shows that the above action
{does not support dS background}. We can easily see that the Minkowski background is 
the only option here {that corresponds to} an exact compensation of the tension
of the initial $D$-brane by the tachyon energy at the bottom of the
potential and the dilaton is a constant.


To produce inflation, in a nearly dS background, we include additional elements in \ref{action_model}.
Such terms may be invoked and expected from several arguments 
\begin{itemize}
	\item Open-closed string interactions in general contain higher vertexes
	beyond the action above. These contributions generate new vertexes
	involving graviton, dilaton and open string tachyon. 
	\item The so called ``marginal deformation'' \cite{Yang:2005ep} excitation
	in the closed strings. This operator is also of a weight zero but
	in fact is non-dynamical at a low-level considerations. However, its
	exclusion by equations of motion will generate additional terms to
	an effective action as well. 
	\item Once a general (not linear) conformal field theory of the dilaton is considered the above analysis would
	not work. New interactions will be generated since the BRST algebra
	of the primary fields will get modified. 
\end{itemize}

We therefore propose a broader (than (\ref{action_model})) action that includes new possible interactions of tachyon of open string and the dilaton of closed string:
\begin{equation}
S=\int d^{4}x\sqrt{-g}\left[\frac{M_{P}^{2}}{2}\left(\Phi^{2}R+4\pd_{\mu}\Phi\pd^{\mu}\Phi\right)-\frac{{T}}{2}\sum_{n=-\infty}^{\infty}\Phi^{n+1}v_{n}\left(\Box,\Tc\right)\right].\label{action_model_new}
\end{equation}
where $R$ is Ricci scalar, ${T}$ is the tension of the D-brane. {Here, the term for $v_0$ is the one appearing in (\ref{action_model}), i.e. $v_0=v(\Box,\,\Tc)+1$. The other terms $v_{n}\left(\Box,\,\Tc\right)$ for $n\neq 1$ would correspond and convey the higher order couplings of the tachyon potential to the dilaton,}
which in general depends on infinite number of d'Alembertian operators
$\left(\Box\right)$ based on the concepts of SFT (cf., Appendix~\ref{AppSFT}). 

{Action (\ref{action_model_new}) is different from (\ref{action_model}) by new terms involving coupling of dilaton and tachyon. Let us stress that we need to establish whether inflation is possible in this framework, keeping the dilaton constant in the vacuum then we will search for constant curvature solutions. This makes irrelevant to consider higher curvature terms. Before proceeding, the appearance of an explicit dilaton potential does not contradict the ``dilaton theorem'' claim, as this was developed in a pure closed string framework. Moreover, results of \cite{DiVecchia:2015oba} {indicate} that the {open-closed SFT coupling} will waive the ``dilaton theorem'' statement. Overall as such, action (\ref{action_model_new}) is a viable attempt to account open-closed strings couplings during the TC process. Explicit computation
	of all such extra terms into the action within pure SFT considerations is
	beyond the scope of our present analysis.}


To support  action (\ref{action_model_new}) as a proposed framework to extract inflationary cosmology, we have to show a constant
curvature (in particular dS) background solution is viable, when the dilaton field
takes a constant value and the open string tachyon condenses to its minimum. Hence, varying (\ref{action_model_new}) with respect
to the metric $g_{\mu\nu}$, $\Tc$ and $\Phi$ we can show that the
following configuration is a solution 
\begin{equation}
\Phi=\Phi_{0}=1,~\Tc=\Tc_{0},~g_{\mu\nu}\text{ is dS with }R=R_{0}=2\frac{{T}}{M_{P}^{2}}\sum_{n}v_{n,0}\,,\label{dssol}
\end{equation}
together with the following relations fulfilled 
\begin{equation}
\sum_{n}v_{n,0}^{\prime}=\sum_{n}v_{n,0}(3-n)=0\,,\label{dssolrel}
\end{equation}
where prime $^{\prime}$ is the derivative with respect to an argument
and the subscript $0$ means that the function is evaluated at $\Tc=\Tc_{0}$.
We note that $\Phi_{0}$ can be any value and is irrelevant as long
as it is finite, so we took $\Phi_{0}=1$ for simplicity. We will
discuss the question of how generic such configurations
(\ref{dssol}), satisfying (\ref{dssolrel}), may arise in SFT in a separate
forthcoming study \cite{Koshelevetal}.
Therefore, our proposed action (\ref{action_model_new})
can support dS solutions (\ref{dssol}).

\subsection{Quadratic variations around de Sitter background}

\label{sftpert}

The quadratic variation of our background action (\ref{action_model_new})
can be written as two parts in the following way

\begin{equation}
\delta^{(2)}S=\delta^{(2)}S_{M_{P}^{2}}+\delta^{(2)}S_{int}\label{d2sfteffmp2}
\end{equation}

The perturbative modes are $\varphi=\delta\Phi$, trace of the metric perturbations
$h$ (we define $\delta g_{\mu\nu}=h_{\mu\nu}$, $h=h_{\mu}^{\mu}$)
and $\tau=\delta\Tc$.
Furthermore, different spins do not mix in the quadratic action i.e.,
tensor modes do not mix with scalar modes. So, the first part
of the quadratic varied action reads
\begin{equation}
\delta^{(2)}S_{M_{P}^{2}}=\int d^{4}x\sqrt{-g}\frac{M_{P}^{2}}{2}\varphi\left(2\Box+3R_{0}\right)\varphi\,.\label{d2sfteffmp2phi}
\end{equation}
where we substituted $h$ from its equation of motion.

The second part, after a Taylor expansion of
the tachyon potential $v\left(\Box,\,\Tc\right)$ around $\Tc=\Tc_{0}$,
reads

\begin{equation}
\delta^{(2)}S_{int}=-\frac{T}{2}\int d^{4}x\sqrt{-g}\sum_{n}\left[(n+1)n\varphi^{2}v_{n,0}+nv'_{n,0}\varphi f(\Box)\tau+\frac{v''_{n,0}}{2}\tau e^{\gamma(\Box)}\tau\right]\,,\label{d2sft-int}
\end{equation}
where we have used (\ref{tachyonvac}). Accounting the fact that the
open string tachyon on its own is not dynamical, the function $\gamma\left(\Box\right)$
in the exponent must be an entire function but the operator $f(\Box)$
may have eigenvalues. Excluding $\tau$ by its equation of motion
is dictated by $\tau=-\frac{\sum_{n}\left(nv'_{n,o}\right)}{\sum_{n}v''_{n,0}}f(\Box)e^{-\gamma(\Box)}\varphi$.
Substituting this back into (\ref{d2sft-int}) yields 
\begin{equation}
\delta^{(2)}S_{int}=\frac{1}{2}\int d^{4}x\sqrt{-g}\varphi\bar{\Fc}(\Box)\varphi\,,\label{d2sfteffintphi}
\end{equation}
where 

\begin{equation}
\bar{\Fc}(\Box)=-T\left[\sum_{n}\left((n+1)nv_{n,0}\right)-\frac{\left(\sum_{n}nv'_{n,0}\right){}^{2}}{2\sum_{n}v''_{n,0}}f(\Box)^{2}e^{-\gamma(\Box)}\right]\,.
\end{equation}

The second order action (\ref{d2sfteffmp2}) with (\ref{d2sfteffmp2phi}) and (\ref{d2sfteffintphi}) represents a non-local scalar (dilaton) degree of freedom around dS background. It is clear from the above expression that higher curvature corrections
are not relevant for us. Indeed, suppose there is a term in the action
like 
$
\sqrt{-g}\Phi^{2}R^{2}
$,
such a term would produce contributions to $h^{2}$ and $\varphi h$
but as long as our background has constant scalar curvature and constant
dilaton field the final effect of such an additional term would be
just renormalization of constants in action (\ref{d2sfteffmp2phi}).
We see that both the spin-0 excitation of the metric and
the dilaton field are combined into one joint scalar mode. Again,
we can show by explicit computation that including other interactions,
like for instance 
$
\sqrt{-g}\Phi^{2}R^{2}w(\Box,\tau)\,,
$
will result in the same net result when all but one scalar fields
can be excluded by equations of motion which finally results in a
single (non-local) scalar excitation.\footnote{We here note that additional contributions to scalar and tensor modes
	can be generated by means of adding the curvature squared corrections,
	like $R_{\mu\nu}^{2}$ or $C^{2}$ where $C$ is the Weyl tensor Moreover,
	following the recent studies performed in \cite{Koshelev:2016xqb,Biswas:2016egy}
	one has to pay special attention in order to maintain unitarity upon
	inclusion of terms which modify the Lagrangian for tensor modes beyond
	the Einstein's gravity. A standard minimal structure like $C^{2}$
	in the action will generate a massive spin-2 ghost (see \cite{Stelle:1976gc}
	for the first comprehensive study of this question). We therefore
	leave the full consideration as an open question.}
We further mention that the open string sector contains only the tachyon, since
higher mass fields have been integrated out, in the course of the
brane decay consideration (cf. Appendix \ref{AppSFT}). 

Altogether, this Appendix provides a possible mechanism to motivate non-local dilaton from string theory. The further theoretical development of this part is deferred for future investigations.


\bibliographystyle{utphys}
\bibliography{References}

\end{document}